# Topological engineering of interfacial optical Tamm states for highly-sensitive near-singular-phase optical detection

Yoichiro Tsurimaki[1], Jonathan K. Tong[1], Victor N. Boriskin[2], Alexander Semenov[3], Mykola I. Ayzatsky[2], Yuri P. Machekhin[4], Gang Chen[1,*] and Svetlana V. Boriskina[1,*]

[1] Department of Mechanical Engineering, Massachusetts Institute of Technology, Cambridge, MA, USA
[2] National Scientific Center 'Kharkiv Institute of Physics and Technology', Kharkiv, Ukraine
[3] National Technical University 'Kharkiv Polytechnic Institute', Kharkiv, Ukraine
[4] Kharkiv National University of Radio Electronics, Kharkiv, Ukraine
*Corresponding authors: sborisk@mit.edu, gchen2@mit.edu

**Abstract:** We developed planar multilayered photonic-plasmonic structures, which support topologically protected optical states on the interface between metal and dielectric materials, known as optical Tamm states. Coupling of incident light to the Tamm states can result in perfect absorption within one of several narrow frequency bands, which is accompanied by a singular behavior of the phase of electromagnetic field. In the case of near-perfect absorptance, very fast local variation of the phase can still be engineered. In this work, we theoretically and experimentally demonstrate how these drastic phase changes can improve sensitivity of optical sensors. A planar Tamm absorber was fabricated and used to demonstrate remote near-singular-phase temperature sensing with an over an order of magnitude improvement in sensor sensitivity and over two orders of magnitude improvement in the figure of merit over the standard approach of measuring shifts of resonant features in the reflectance spectra of the same absorber. Our experimentally demonstrated phase-to-amplitude detection sensitivity improvement nearly doubles that of state-of-the-art nano-patterned plasmonic singular-phase detectors, with further improvements possible *via* more precise fabrication. Tamm perfect absorbers form the basis for robust planar sensing platforms with tunable spectral characteristics, which do not rely on low-throughput nano-patterning techniques.

Optical transduction is a widely used detection mechanism in remote sensing and monitoring of a variety of physical, chemical, and biological events. It is based on measuring environmental changes by detecting the change in one of the characteristics of light interacting with the target medium, including its amplitude, wavelength, incident angle, and phase. Optical sensing is intrinsically non-invasive, and can be used in extreme conditions, such as high toxicity, high temperatures, electrical noises, or strong magnetic fields. Among many types of optical sensors, surface-plasmon polariton (SPP) sensors are widely investigated and used[1–7]. Evanescent fields of SPP modes supported by metallic structures strongly interact with the surrounding dielectric medium, and environmentally-induced changes of their propagation constants provide an optical transduction mechanism[4]. Excitation of localized SPP modes generates strong electromagnetic field in a small volume close to the metal-dielectric interface, making possible detection of very small variations of the local refractive index. Hence, the SPP sensors offer highly-sensitive, label-free, and non-destructive optical detection and monitoring of chemical and biological reactions on the surface. A common scheme of the SPP excitation on planar interfaces is the Kretschmann−Raether scheme, where a prism is placed on top of the metal film to excite surface plasmon polariton (SPP) waves by incident light. While the use of prism is required to achieve coupling of propagating light to high-momentum SPP optical states, it hinders applications of the Kretschmann-Raether scheme to multiplexed bio/chemical sensors composed of arrays of multiple sensing elements[8].

In order to overcome this problem, sensors based on excitation of localized surface plasmon (LSP) resonances on nanoscale metallic structures and arrays have been demonstrated[1–3,8–14]. The LSP resonant frequency depends on the geometry of nanostructure, and thus a wide variety of structures has been investigated to improve the detection sensitivity. However, fabrication of plasmonic nanostructures often requires use of low-throughput complicated fabrication processes. Furthermore, the effective area used for strong light-matter interactions is small, as the total area of the metal nanostructures is often only a fraction of the sensor substrate area, which reduces the optical signal level of LSP-based sensors compared to that of SPP-based ones. Therefore, a sensing platform that does not rely on nano-patterning, is easy to fabricate, is accessible without external bulky couplers, and exploits the whole surface area of the sensor to maximize the detection sensitivity is of great interest.

When the amplitude of light reflected from the sensor is exactly zero, the phase of light becomes undefined, i.e., singular[15–18]. As the frequency is scanned through the point where the perfect absorption is observed, the phase of the reflected light exhibits an abrupt jump at this singular point. Even in the case of near-zero reflection, the phase exhibits rapid variations in the frequency range around reflection minimum. This offers opportunities for optical sensing using optical phase rather than amplitude variations as the transduction mechanism[19,20]. Phase-based detection schemes have already been demonstrated to achieve higher sensitivity than the conventional amplitude-based ones[3,19,21–24]. Recently, plasmonic nanostructured sensors exhibiting near-perfect absorption at the resonance condition showed sensitivity high enough to detect a single molecule when operated in the singular-phase sensing regime[19,22]. Despite high sensitivity, these sensors suffer from the same drawbacks as the conventional LSP amplitude-based detectors, including complicated low-throughput lithographic fabrication and the reduction of the active sensing region to a fraction of the surface area. To simplify the fabrication process of the singular-phase sensors, new types of plasmonic perfect absorbers prepared by the methods of colloidal chemistry have been recently explored. These include self-assembled three-dimensional metamaterials fabricated from silver nanoparticles covered with a silica shells[25] and more complex lamellar metamaterials consisting of stacked polymer layers doped with gold nanoparticles alternating with undoped polymer layers[26,27]. However, this fabrication approach is restrictive in the spectral positioning of resonant features, and still requires several fabrication steps. Therefore, simple-to-fabricate perfect absorber structures that show high sensitivity and high signal level owing to the strong light-matter interaction region spanning the whole area of the sample are of great interest for practical applications.

Localized optical states other than the SPP or LSP modes can be supported at interfaces between two dissimilar optical materials, with the amplitude of the electric field decaying exponentially away from the interface. The existence of these interfacial states is guaranteed and protected by topological properties of the materials forming the interface, namely, by the inversion of the geometrical (i.e., Berry) phase across the interface[28]. Interfacial optical states can exist between dissimilar one- (1D) and two-dimensional (2D) dielectric photonic crystals (PhCs)[28,29] and at an interface between a 1D photonic crystal and a noble metal[30–33]. These states are topologically protected by the inversion of the Zak phase across the interface. Zak phase is the Berry phase acquired during the adiabatic motion along an energy band across the Brillouin zone of a material[34], which can be used as an invariant characterizing topological properties of the bands of planar multi-layered structures. The optical state supported by the metal-PhC interface is commonly referred to as the Tamm plasmon state, in analogy to interfacial localized electron states predicted by Russian physicist Igor Tamm[35]. Energies of optical Tamm states lie within the forbidden energy gap of the photonic crystal and exhibit parabolic dispersion with respect to the in-plane wave-vector (i.e., the angle of incidence), with slightly different 'effective masses' for the two orthogonal polarizations of electromagnetic field[30,32]. Because the parabolic dispersion curves of Tamm states are within the light cone, they can be excited by propagating waves at either normal or oblique incidence without the use of a prism or gratings. The nanophotonics structures supporting optical Tamm states include simple planar multi-layered films, which do not require complicated fabrication. They can be easily deposited on a variety of semiconductor and metal surfaces, and their resonant frequency is highly tunable by design[30,32]. Coupling of incident light to the optical Tamm state can be optimized *via* structure design to result in perfect absorption (i.e., zero reflection) at a given frequency[36]. These attractive characteristics of the optical Tamm structures already led to a variety of their applications in optoelectronics and photonics, including amplitude-based optical sensing[32,37,38], development of metal/semiconductor lasers[39–41], etc.[42–49].

In this paper, we theoretically and experimentally demonstrate that perfect absorbers based on the excitation of optical interfacial Tamm states can be highly promising platforms for robust and sensitive optical sensing. The fabricated Tamm sensors are planar metal-dielectric multilayered structures produced by e-beam deposition and plasma-enhanced chemical vapor deposition (PECVD), which can be easily tunable to operate at a specific wavelength across the visible and the infrared spectral ranges. Our results illustrate how the use of the singular-phase detection scheme can provide a 37-fold improvement of the sensitivity of Tamm sensors over the conventional amplitude-based scheme. Furthermore, the sensor figure of merit, which takes into account the quality factor (i.e., linewidth) of the resonant feature as well as its sensitivity to the environmental changes, is improved by a factor of 481 by the use of the singular-phase detection scheme.

## Tamm structures design

We developed Tamm structures comprised of 1D finite-size dielectric PhCs on top of thin metal films deposited on dielectric substrates (Fig. 1a). In order to achieve zero-reflection (i.e., singular-phase) condition useful for sensing, the structures were optimized to provide perfect narrowband absorption at a specific wavelength and angle of incidence *via* excitation of the optical Tamm states on the PhC-metal interface. This was accomplished by tuning the surface impedance of the bottom surface of the PhC facing the metal to match the impedance of the metal film on a substrate. It has been previously proven that the interfacial state is guaranteed to exist between two lossless and perfectly reflective media with the inverse purely reactive surface impedances: $Z_1 = iz'' = -Z_2$[28]. The existence of the Tamm state is topologically protected under such conditions because the optical impedance of a surface is directly related to the topological properties of the bulk material through the geometrical (Zak) phases[34] of its photonic bands.

For surfaces of real materials – such as short sections of PhCs and thin films of real metals used in this work – optical impedance is a complex number. If the metal-PhC interface is located at $z = z_0$, optical impedance of the 1D PhC surface can be defined as $Z_{PhC} = E_x(z = z_0 + 0^+)/H_y(z = z_0 + 0^+)$ and that of the metal layer on a substrate as $Z_{metal} = E_x(z = z_0 + 0^-)/H_y(z = z_0 + 0^-)$. $E_x$ and $H_y$ are the total electric and magnetic fields on the PhC and metal surfaces, respectively (Fig. 1a). To achieve maximum power transfer from the free space propagating wave into the Tamm state, the complex optical impedance of the metal layer has to be conjugate-matched to that of the bottom surface of the PhC, $Z_{PhC} = Z_{metal}^* = z' + iz''$. In turn, the top surface of the Tamm absorber needs to provide conjugate impedance matching to the free space, where $Z_0 = \sqrt{\mu_0/\varepsilon_0}$ is the vacuum impedance, and $\mu_0$, $\varepsilon_0$ are the vacuum permeability and permittivity, respectively. Conjugate impedance matching is a technique routinely used in the transmission lines design to maximize power transfer from the source to the load [50], which can also be applied to design nanoantennas and super-absorbers[51,52]. Note that in the case of a purely reactive load with an imaginary surface impedance (e.g., an infinite lossless PhC), the conjugate matching condition takes the form of the impedance inversion[28], while for the case of a purely real source impedance (e.g., the vacuum impedance) it reduces to the conventional impedance matching condition. Metal surfaces at frequencies below their plasma frequency have impedances that are predominantly reactive (i.e., with $z'' > z'$), which makes them highly reflective due to strong impedance mismatch with the free space. However, the conjugate impedance matching technique can help transform a near-perfectly reflective surface to a perfectly absorbing one at a given frequency at which all the plane wave energy is coupled to the interfacial Tamm state.

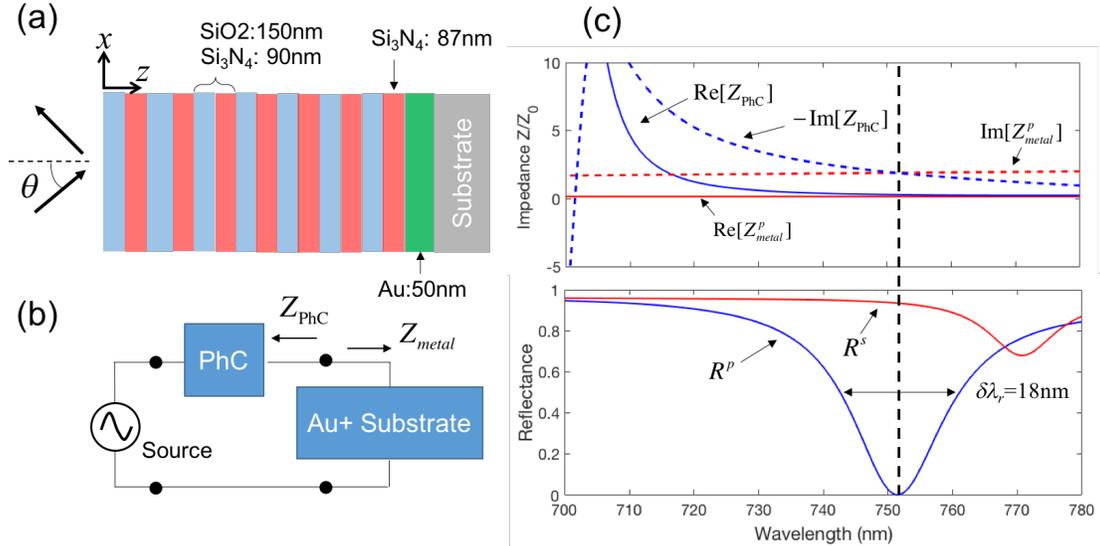

**Fig. 1. Tamm perfect absorber design**. (a) Planar Tamm absorber structure composed of 7 periods of a 1D dielectric photonic crystal and a thin gold film on a dielectric substrate. The light is incident on the structure at an oblique angle $\theta$ measured from the normal to the surface. (b) An equivalent conjugate-matching circuit. (c) Optical impedances $Z_{PhC}^*$ and $Z_{metal}$ for p-polarization as a function of the wavelength (top panel) and s-and p-polarized reflectance spectra of the Tamm structure (bottom panel).

We designed a planar Tamm absorber composed of a finite-length 1D photonic crystal made of alternating thin layers of low-index silica ($SiO_2$) and high-index silicon nitride ($Si_3N_4$) on top of a thin gold (Au) film deposited on an optically-thick amorphous Si substrate as shown in Fig. 1a. The thicknesses of the layers as well as the number of periods in the PhC were optimized to conjugate-match the 1D PhC surface to that of the Au film and also conjugate-match the whole structure to the free-space. Complex optical impedance of a surface is directly related to the complex surface reflection coefficient $r = |r|e^{i\delta}$ as $Z/Z_0 = (1 + r)/(1 - r)$, and can be easily calculated with the rigorous semi-analytical transfer matrix method (see Methods). This provides a quick and low-numerical-cost approach to designing structure with interfacial states as opposed to time- and memory-hungry numerical search for complex eigenvalues of these interfacial states. The equivalent conjugate-matched transmission line schematic of the Tamm structure is shown in Fig. 1b.

In the simulated structure, the thickness of the gold film was set to 50nm in order to use the thin-film optical properties of Au measured by Reddy et al.[53] (supplementary Fig. S1). The incident angle of the plane wave was set to 60° to normal to match experimental conditions of ellipsometric measurements used to characterize the singular-phase Tamm sensors (see Methods). Numerical modeling of the conjugate-matched 1D PhC yielded an optimized structure design consisting of 7 periods of the photonic crystal with the layer thicknesses of 150 and 90nm for $SiO_2$ and $Si_3N_4$, respectively. The thickness of the bottom $Si_3N_4$ layer of the PhC was tuned to measure 87nm, slightly different from the other $Si_3N_4$ layers, to achieve the conjugate-matching condition on the PhC-Au interface. Optical properties of $SiO_2$, $Si_3N_4$, and amorphous Si were taken from Palik[54] (supplementary Fig. S1). The upper panel of Fig. 1c shows the calculated real and imaginary parts of the optical impedances $Z_{metal}$ and $Z^*_{PhC}$ for the p-polarized incident light. As seen in the figure, the imaginary parts of optical impedances of the two surfaces facing each other take opposite values at one wavelength, with the real parts approaching each other at the same wavelength. This near-conjugate-matching condition creates and protects the optical Tamm state at the PhC-Au interface. This state is not perfectly confined and is characterized by both radiative and dissipative losses. The increase in the number of periods of the PhC reduces the radiative loss, while increases the non-radiative loss due to residual parasitic absorption in dielectric materials.

The calculated reflectance of the metal-dielectric planar Tamm absorber comprised of the conjugate-matched 1D PhC and an Au film for s- and p-polarized light is plotted in the lower panel of Fig. 1c. The structure exhibits near-perfect absorption (reflectance of $10^{-5}$) of p-polarized plane waves at wavelength of 751nm and incident angle of 60 degrees to normal. This feature in the reflectance spectrum is a manifestation of the efficient coupling, i.e. perfect absorption, of the incident light to the optical Tamm state on the PhC-Au interface at the frequency where conjugate-matching is achieved. The optical Tamm state resonant wavelength is located in the first photonic bandgap of the PhC (supplementary Figs. S2a,b). The structure is thus highly reflective at frequencies around the absorption maximum, resulting in formation of the high-contrast narrowband spectral feature favorable for sensing applications. Even though in this configuration the perfect absorption only occurs for the p-polarized incident light, weak coupling to the optical Tamm state is also observed at around 771nm for s-polarized light (Fig. 1c). The parabolic dispersion characteristics of both the p- and s-polarized optical Tamm states are shown in supplementary Figs. S2c,d. It should also be noted that an alternative way of observing the resonant features in the optical spectrum of Tamm structures is based on fixing the wavelength of the incident light and monitoring the changes of the reflectance as a function of the angle of incidence (Supplementary Fig. S3).

Figure 2 shows the spatial distribution of the absorbed power density $P_{abs}$ of the *p*-polarized wave in the three bottom layers of the PhC as well as in the gold film at the resonant wavelength (751.5nm) and a nearby off-resonant wavelength (720nm). The absorbed power density is normalized by the power density under the assumption that the incident power is completely absorbed in the Au film $P_0$ and distributed uniformly within the film. $P_0$ is introduced as a reference absorbed power density, which is estimated by the total incident power used in the simulation divided by the volume of the Au film. Normalization procedure allows plotting the on- and off-resonance power density distributions independently of the incident power and on the same scale for comparison. It can be seen that at the resonant wavelength, the absorbed power density is tightly concentrated near the PhC-Au interface supporting the optical Tamm state. At the off-resonant wavelength, absorbed power is negligibly small, as the interaction between the incident light and the optical Tamm state is weak.

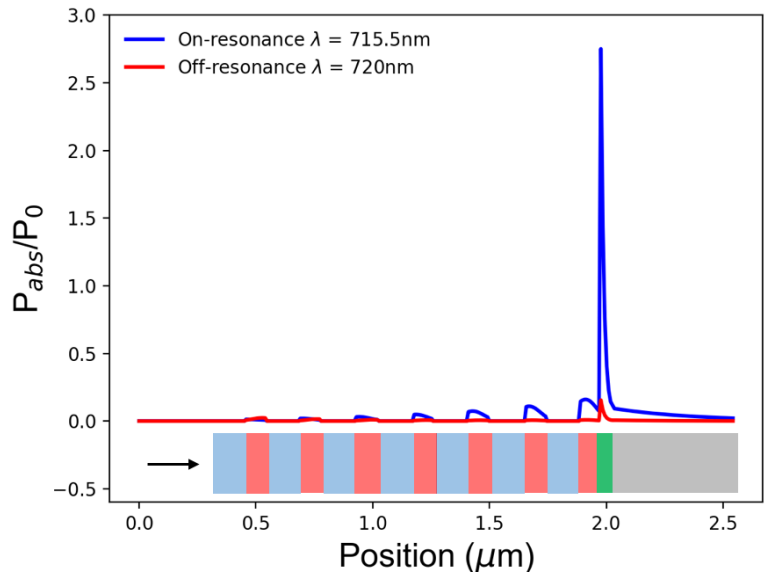

**Fig. 2. Absorbed power density near the interface that supports the optical Tamm state**. Absorbed power density normalized to the power density under the condition that all the incident power is uniformly absorbed in the Au layer at on-resonance wavelength of 751.5nm (blue line) and off-resonance wavelength of 720nm (red line). The inset shows the schematic of the Tamm structure and the direction of the light incidence.

Reflection of wave from interfaces is typically accompanied by the phase shifts of reflected light. Linearly polarized light incident at an oblique angle *θ* becomes elliptically polarized upon reflection from the planar Tamm structures. The change of magnitude and phase for two polarized light components can be quantified by two ellipsometric angles $\Psi = \text{atan}(|r_p|/|r_s|)$ and $\Delta = \delta_p - \delta_s$. Here, $r_{p,s} = |r_{p,s}|e^{i\delta_{p,s}}$ and $\delta_{p,s}$ are the complex reflection coefficients of the Tamm absorber and the phases for *p*- and *s*-polarized light, respectively. Figure 3 shows the ellipsometric angles of the structure shown in Fig. 1a as a function of the wavelength. The ellipsometric angle $\Psi$ is plotted as the blue solid line in Fig. 3a and reaches the minimum at the same wavelength as the reflectance (see Fig. 1c)[19]. This resonant feature in the amplitude ellipsometric angle spectrum is accompanied by a jump of the phase ellipsometric angle $\Delta$ by the amount of $\pi$ at the wavelength of the Tamm resonance (Fig. 3b). Here, the range of $\Delta$ values is chosen to vary from -90 to +270 degrees, as this is the operational range of ellipsometer we use in the phase measurements. Note that the rapid variation of $\Delta$ by the amount of $\pi$ occurs between $\Delta = 240°$ and $\Delta = 90°$, while due to the cyclic nature of the phase, $\Delta = -90° \equiv 270°$, and the curve does not in fact exhibit a discontinuity. Comparison of the linewidth of the reflectance resonance ($\delta\lambda_r = 18$nm) to the linewidths of the spectral features of the ellipsometric angles ($\delta\lambda_\Psi = 8.3$nm and $\delta\lambda_\Delta = 0.006$nm) shows that measurements of the phase response of the Tamm sensor can contribute to improvements of its detection limit.

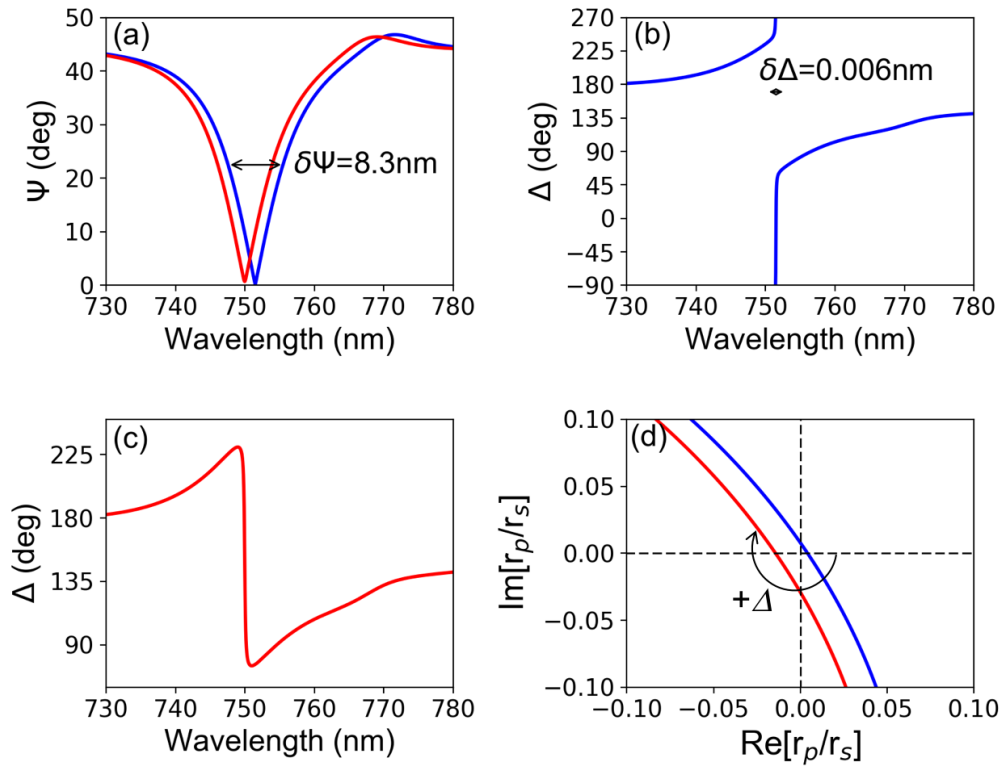

**Fig. 3. Ellipsometric angles $\Psi$ and $\Delta$ in the vicinity of the singular point in the Tamm absorber spectrum.** (a) Ellipsometric angle $\Psi$ of the structure in Fig. 1a as a function of wavelength around the Tamm resonance (blue line). The red line shows the same spectrum for a Tamm structure with the geometry slightly deviating from the optimized one shown in Fig. 1 (here, the bottom $Si_3N_4$ layer thickness is 86nm). (b,c) The corresponding ellipsometric angles $\Delta$ for the structures with parameters as in panel a, (bottom nitride layer thickness is 87nm in (b), blue line, and 86nm in (c), red line). (d) Evolution of the ellipsometric angle $\Psi$ values in the complex plane as the wavelength is scanned through its resonant value. The optimized structure curve passes through the first quadrant (blue line), while the curve for the slightly modified structure passes through the third quadrant (red line).

As the phase experiences drastic changes in the vicinity of the singular point in the Tamm structure spectrum, slight fabrication-induced variations of the absorber geometry may result in significant variations of the shape of the asymmetric feature in the phase spectrum. However, the phase jump by $\pi$ should still be observed, and can be used for high-sensitivity measurements. Figures 3a and 3c show the ellipsometric angle spectra of the Tamm absorber with the geometry slightly detuned from the optimized one (red lines in Figs. 3a,c). The rapid phase variation occurs at the near-singular point, but the shape of the curve in Fig. 3c is different from that in Fig. 3b. Figure 3d illustrates the origin of this phenomenon, by plotting the real and imaginary parts of the amplitude ellipsometric angle in the complex plane. When this curve shifts to another quadrant due to the small change of parameters, the shape of the phase spectrum changes. Regardless of the exact shape of this asymmetric spectral feature, it provides a powerful mechanism of sensor performance improvement via sensitivity enhancement. To make use of this performance improvement, we used spectroscopic ellipsometry (see Methods) to demonstrate singular-phase sensing with Tamm structures. The procedure used for the calculations of the spectral widths of symmetric and asymmetric resonant features is shown in supplementary Fig. S4.

## Non-contact temperature detection with Tamm absorbers

To quantify experimentally the performance of the Tamm structure as optical sensors in both the amplitude and the phase detection regimes, we fabricated the planar absorber with optimized parameters shown in Fig. 1a. Gold and dielectric layers have been deposited by e-beam deposition and plasma-enhanced chemical vapor deposition. Figure 4a shows a SEM image of one of the fabricated samples. Optical reflectance spectra of the fabricated structure have been measured with a spectral ellipsometer. A schematic of the ellipsometric measurements procedure is shown in Fig. 4b. The samples were mounted on top of a home-built temperature control system to achieve temperature tuning, and their spectral properties were characterized in the ambient conditions without humidity control. The detailed information of the fabrication and the measurements is given in Methods. Figures 4c-e show the measured spectral reflectance of the sample for $s$- and $p$-polarized incident light, and the corresponding spectra of the ellipsometric angles Ψ and Δ, respectively. The incident angle $\theta$ was around 60º and the temperature of the sample was fixed at 25ºC, which was monitored by a K-type thermocouple. The circles represent the experimental data measured by the ellipsometer at discrete wavelength points.

Reflectance spectra in Fig. 4c exhibit resonant reflectance dips at around 766.4nm and 744.2nm for $s$- and $p$-polarized incident light, in reasonable agreement with the predictions of the numerical simulations (771nm and 751.5nm, respectively). The corresponding on-resonance minimum reflectance values were measured as 0.52 and 0.10, respectively. The deviation of the experimental spectra from the numerical calculations is likely due to the imperfections in thicknesses, surface roughness and the optical properties of the fabricated sample. The measured ellipsometric angle Ψ reaches the minimum value of around 1.8 at 744.2nm, while the ellipsometric angle Δ exhibits a drastic change at this wavelength by the amount of 180º within the wavelength range of less than 3nm (Fig. 4d,e). At off-resonance wavelengths, the ellipsometric angle Ψ measures around 45º because the amplitudes of the reflection coefficients for $s$- and $p$-polarized waves are comparable. We would like to emphasize that although the complete darkness condition has not been observed in the reflectance spectrum of the absorber due to fabrication imperfections, the Tamm state resonance and the corresponding near-singular phase behavior can be clearly seen in Figs. 4c-e. The topological protection of the Tamm interfacial state formation makes it very robust to fabrication errors, allowing use of less precise high-throughput fabrication techniques.

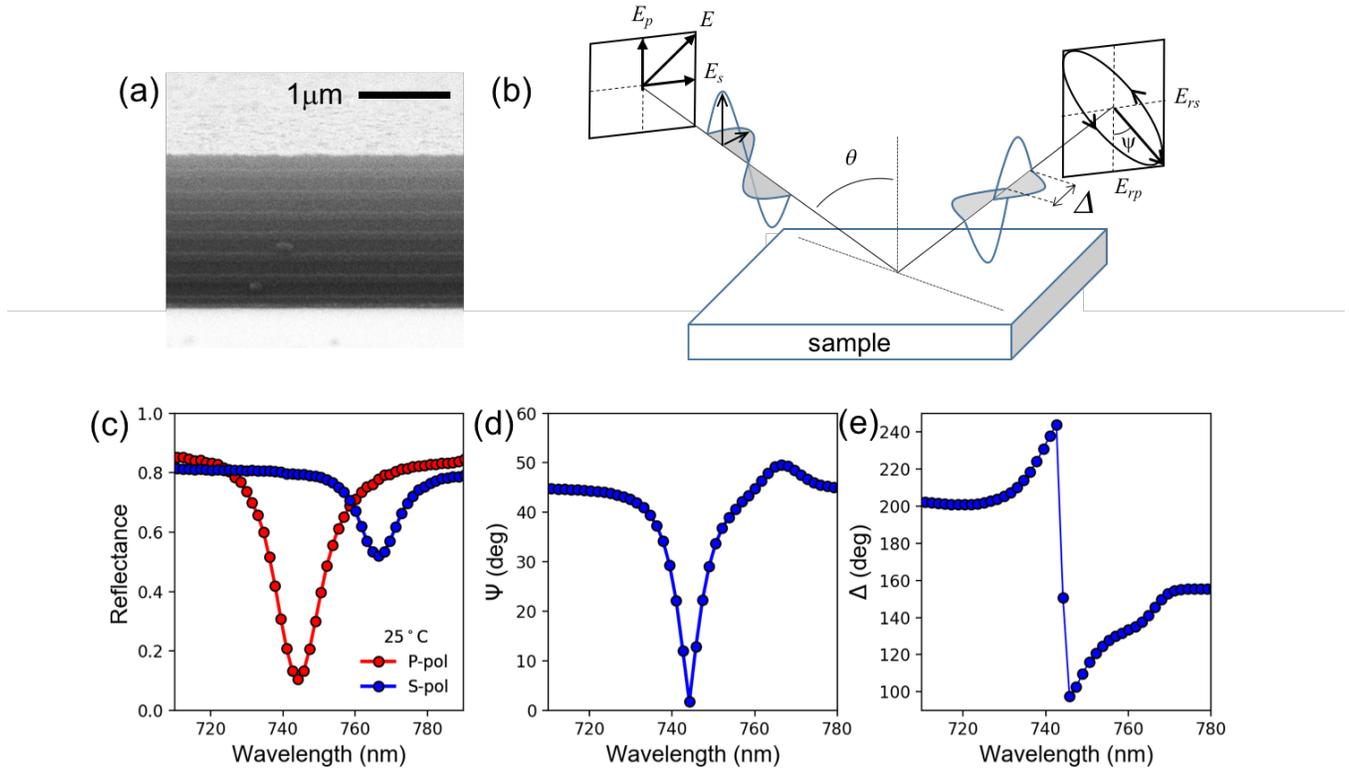

**Fig. 4. Room-temperature spectral characterization of the Tamm absorber.** (a) SEM image of one of the fabricated Tamm absorber samples. (b) A schematic of the ellipsometric measurements that provide both amplitude and phase spectral characteristics of the sample. (c) Measured reflectance of the sample in panel (a) for *s*- and *p*-polarized light incident at an angle of 60 degrees to normal at room temperature (25°C). (d) Ellipsometric angle Ψ as a function of the wavelength measured under the same conditions. (e) The corresponding ellipsometric angle Δ measured as a function of the wavelength.

We tested the performance of the fabricated Tamm structure as the non-invasive label-free optical temperature sensor. Temperature detection *via* optical transduction does not require a physical contact with the sample and is based on sensing small changes in the refractive index and to a lesser extent in the thicknesses of material layers, both of which typically increase with temperature, resulting in the amplitude and phase changes of the reflected light[55–58]. We experimentally investigated and compared three optical sensing methods: the spectral sensing, in which the shift of the reflectance minimum is monitored at a fixed angle, the amplitude sensing, in which the change of the amplitude of reflectance is monitored at fixed wavelength and angle, and the phase sensing, in which the change of phase of light is monitored at fixed wavelength and angle. Figure 5a shows the change in the spectral reflectance of *p*-polarized light from the sample in response to the temperature variation from 25 to 32°C. Figure 5b shows the same data in the wavelength range just around the resonant wavelength. The tight overlap of the spectra for three different temperatures as well as the limited (1.5nm) resolution of the ellipsometer does not allow for the reliable measurements of the spectral shift of the resonant feature. To estimate this shift from the measured data, we linearly extrapolated the reflectance measurement points near the resonant minima, and found that the shift was 0.04nm when the temperature increased from 25°C to 32°C (Fig. 5b). Thus, the detection sensitivity estimated from the data in Figs. 5a,b equals to $S_r = d\lambda_{res}/dT = 0.0057(nm/\,^oC)$, where $d\lambda_{res}$ is the shift of the resonant wavelength with the change of the temperature. The spectral width $\delta\lambda_r$ of the resonant feature in the reflectance spectrum is 15nm (the corresponding Q-factor is Q~50). The standard figure of merit of an optical sensor is defined as a ratio of the sensitivity and the linewidth[59,60], which yields the figure of merit for the amplitude detection scheme $FOM_r = S_r/\delta\lambda_r = 3.8\cdot10^{-4}\,^oC^{-1}$.

The changes in the temperature cause thermal expansion of the layers comprising the Tamm structure, as well as the change in their refractive indices. Thermo-optic coefficients of $SiO_2$ and $Si_3N_4$ are very small, and the thermal expansion coefficients of materials are also negligible – $0.6\cdot10^{-6}$ K$^{-1}$, $3\cdot10^{-6}$K$^{-1}$, and $14\cdot10^{-6}$K$^{-1}$ for $SiO_2$, $Si_3N_4$, and Au, respectively. Therefore, the changes in the optical spectrum of the Tamm absorber primarily result from the thermally-induced modification of the Au refractive index and the amorphous silicon substrate, and are not pronounced. Therefore, spectral sensing would require using a sensitive optical measurement systems as well as signal processing methods to provide robust and reliable detection of spectral shifts on the order of $10^{-2}$ nm.

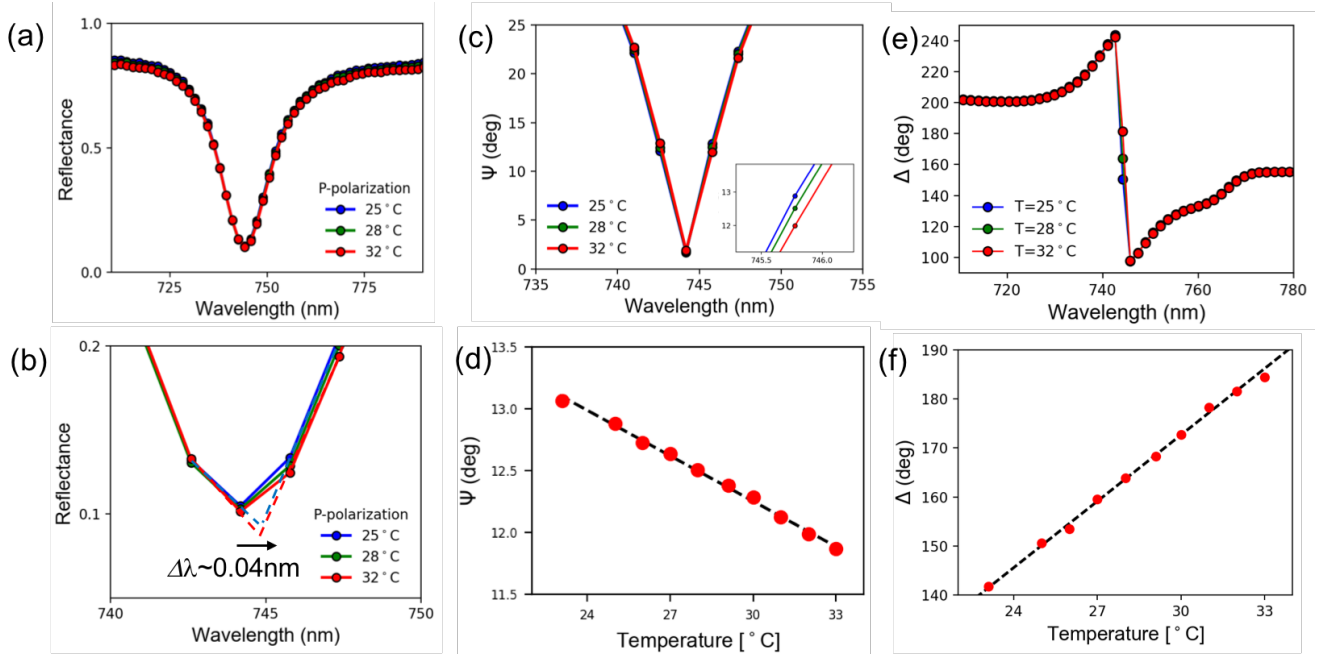

**Fig. 5. Comparison of spectral, amplitude, and singular-phase temperature detection schemes**. (a) Reflectance of *p*-polarized light from the Tamm sensor for three different temperatures of the sample. (b) A close-up view of this same data as is panel (a) in the vicinity of the Tamm resonance. (c) Amplitude ellipsometric angle Ψ as a function of wavelength for three different temperatures. (d) Ellipsometric angle Ψ as a function of the temperature at a fixed wavelength of 745.8nm. (e) Phase ellipsometric angle Δ as a function of the wavelength for three different temperatures. (f) Ellipsometric angle Δ as a function of the temperature at a fixed wavelength of 744.2nm.

The detector sensitivity can however be increased by using alternative detection schemes. Figure 5c shows the measured spectra of the amplitude ellipsometric angle Ψ at three different temperatures. In our measurements, the temperature-induced maximum change of the ellipsometric angle was observed at wavelengths in the immediate vicinity of the minimum in the Ψ spectrum. The inset of Fig. 5c shows the close-up view of the Ψ spectra around wavelength of 745.8nm. In Fig. 5d, we plot the ellipsometric angle Ψ as a function of temperature around the ambient value at the same wavelength, which shows linear change with temperature. The slope of the change represents the sensitivity, and was estimated from the measured data as $S_\Psi = d\Psi/dT = 0.12 \deg/\,^oC$. The spectral width $\delta\lambda_\Psi$ of the resonant feature in the amplitude ellipsometric angle spectrum is 6.7nm (the corresponding Q-factor is Q~107), which yields the figure of merit for the amplitude detection scheme as $FOM_\Psi = S_\Psi/\delta\lambda_\Psi = 0.018 \deg/(\,^oC \cdot nm)$.

Finally, Figs. 5e-f show the results of the spectral measurement using the singular-phase detection approach. The phase ellipsometric angle Δ spectra acquired at three different temperatures of the sample are shown in Fig. 5e. The ellipsometric angle Δ exhibits a change by the amount close to 180° within the three measurement points (i.e. within the 3nm narrow spectral bandwidth). Under the zero-reflection condition, a discontinuous jump of the Δ value by 180° (π) is expected. However, our sample does not exhibit the complete darkness at the resonant wavelength due to imperfections in fabrication. As a result, the ellipsometric angle Δ varies by slightly less than 180°, and this variation is continuous. Nevertheless, at the resonant wavelength of 744.2nm, the ellipsometric angle Δ shows a large variation as temperature of the sample varies. Figure 5f shows how the ellipsometric angle Δ scales as a function of temperature at the fixed wavelength of 744.2nm. Similar to the amplitude ellipsometric angle, the phase ellipsometric angle exhibits linear change as a function of temperature. The sensitivity calculated from measuring the slope of the plot in Fig. 5f was estimated $S_\Delta = d\Delta/dT = 4.5(\deg/\,^oC)$. The spectral width of the experimentally measured ellipsometric angle Δ is hard to determine due to the drastic change of Δ within the wavelength range of 3nm and the spectral resolution of 1.5nm of the ellipsometer. Although it is obvious that the spectral width is less than 3nm, a more precise determination of the spectral width by curve fitting entails uncertainty. Therefore, we determined the spectral width of the ellipsometric angle Δ by simulating the same system that exhibits the same minimum $\Psi_{min}$~1.8 at the same wavelength of resonance at 744.2nm (See supplementary Fig. S4). As a result, we found that the spectral width $\delta\lambda_\Delta$ of the resonant feature in the phase ellipsometric angle spectrum was 0.52nm. Therefore, the overall figure of merit for the phase sensing scheme is estimated to be $FOM_\Delta = S_\Delta/\delta\lambda_\Delta = 8.65(\deg/\,^oC \cdot nm)$.

## Discussion

Our results demonstrate that the near-singular-phase sensing with the planar Tamm near-perfect absorber was the most sensitive method among the three detection schemes tested. The experimental sensitivity improvement of the phase detection scheme over that of the amplitude detection under the same intensity of light collected by the detector is $S_\Delta/S_\Psi = 37$. This is almost double the sensitivity enhancement experimentally demonstrated with the singular-phase sensor based on a nanopatterned array of Au nanoantennas (~20)[19], and four times the value of the enhancement predicted for the singular-phase sensor based on excitation of SPP wave on a planar Au surface (~9) [22]. It should be noted that the improvement in the figure of merit over the singular-phase detection scheme over the amplitude one is even more dramatic, $FOM_\Delta/FOM_\Psi = 481$, owing to the Q-factor increase of the resonant features in addition to the sensitivity improvement. The simple one-dimensional geometry of planar Tamm sensors in combination with the high sensor figure of merit makes them an attractive choice for a singular-phase detection platform.

The sensitivity enhancement of the Tamm singular-phase sensor can be much larger if the total darkness condition is realized. It has been previously shown that the sensitivity improvement of the singular-phase detection scheme over that of the amplitude one scales as $S_\Delta/S_\Psi \sim 1/\Psi_{min}$, where $\Psi_{min} = \Psi(\lambda_{res})$ is the value of the amplitude ellipsometric angle at the frequency of the absorption resonance measured in radians[3,19]. Accordingly, as $\Psi_{min}$ tends to zero in the case of zero reflectance resonance, higher sensitivity can be achieved. Our experimental data support this trend, as can be seen in Fig. 6. The red circles on this plot correspond to the sensitivity improvement $S_\Delta/S_\Psi$ calculated from the experimentally measured ellipsometric angle $\Delta$ and the corresponding minimum of the ellipsometric angle $\Psi_{min}$ data taken at different angles of excitation $\theta$. As $\theta$ deviates from the optimum value of 60 degrees, the Tamm reflectance resonance becomes less deep, and the $\Psi_{min}$ value increases accordingly. The measured data taken at different $\theta$ fall onto the theoretical curve of $1/\Psi_{min}$, which illustrates that significant sensitivity enhancement is possible with the samples exhibiting deeper resonances approaching the complete darkness condition.

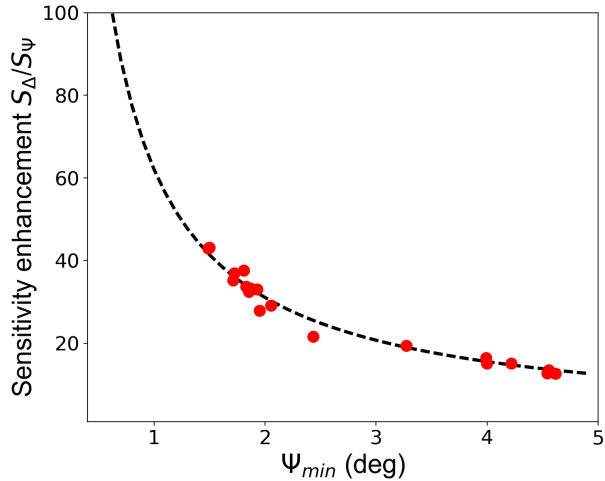

**Fig. 6. Phase sensitivity improvement as the complete darkness condition is approached**. Experimental sensitivity of the singular-phase detection scheme scaling with the minimum value of the amplitude ellipsometric angle at the Tamm resonance wavelength. Red dots are the measured results, and the dashed line is the trend line that scales as $1/\Psi_{min}$.

The sensor detection limit – i.e., the minimum temperature change that can be detected by a sensor – depends not only on the sensor sensitivity but also on the resolution of the instrument used for measurements of the temperature-induced changes. The spectral resolution of the ellipsometer we used in our measurements is 1.5nm. As the shift of the resonant peak is on the order of $10^{-2}$nm/K, it is not realistic to use the spectral sensing scheme in our ellipsometer, which stems from both the low sensitivity and the resolution of the ellipsometer. The ellipsometer has the uncertainty in the measurements of the amplitude ellipsometric angle of around $\delta\Psi=\pm0.033$ degrees at the resonant wavelength. Consequently, we concluded that the minimum temperature variation detectable in the current measurement system is $0.28^oC$ for the amplitude detection scheme. In turn, the uncertainty in the measurements of the phase ellipsometric angles is around $\delta\Delta=\pm0.99^o$ at the resonant wavelength, and thus the minimum temperature variation detectable in the current measurement system was $0.2^oC$ for the phase detection scheme. The signal detection scheme that attains low phase noises, i.e. small errors of measurement of $\delta\Delta$, will significantly improve the detection limit. In general, the minimum detectable angle of the ellipsometric angle $\Delta$ under properly constructed signal detection scheme is on the order of $10^{-3}$-$10^{-2}$ degrees[22]. Therefore, by using the low noise signal detection scheme, the minimum temperature difference that can be measured with the singular-phase detection can be on the order of $10^{-4}$ $^oC$ under the measured sensitivity of $S_\Delta = 4.5(deg/$ $^oC)$ of our sample, and even smaller under the near-darkness condition (see Fig. 6). Therefore, the singular-phase detection with the optical Tamm state can provide highly-sensitive tunable optical sensors that are simple to design and fabricate and easy to scale up.

Overall, non-invasive interferometric temperature monitoring is a very useful alternative to conventional thermocouple or pyrometer measurements and finds use in remote monitoring of fabrication processes such as molecular beam epitaxy, reactive ion etching, and rapid thermal processing[55–58]. It becomes invaluable in the situations where the sample needs to be insulated from the environment (e.g., in a vacuum chamber [57]), and at ultra-low and ultra-high (up to $1600^oC$ [58]) temperatures. It also significantly outperforms thermocouple-based measurements in the presence of strong magnetic and varying electromagnetic fields (where thermocouple temperature errors up to 1K have been reported[56]), and in the situations when the temperature changes rapidly due to slow temporal response of thermocouples [56]. Typical interferometric temperature sensors that make use of the temperature-induced refractive index change and operate in the spectral detection regime require much longer optical paths (on the order of hundreds of micron) within the material to compensate for very small thermo-optical coefficients of most materials[57]. It also requires the sensor material to be optically transparent. This makes the singular-phase Tamm sensors – which only require a thickness of 1-2 micron and can use high-thermal-conductivity metals as absorbers – especially attractive platforms for sensitive detection with high temporal resolution, which can be applied to any surface whose temperature needs to be monitored.

Unlike nano-patterned sensors using metal nanoantennas, planar Tamm sensors can be used in harsh environmental conditions, including corrosive atmosphere and high temperatures, which can cause severe degradation and melting of nanoantennas. Planar structures that support the optical Tamm state can be made of a variety of materials for different purposes and different conditions of operation. Gold can be easily replaced with either low-loss materials such as silver (supplementary Fig. S5) or cheaper materials such as aluminum (supplementary Fig. S6), as well as other metals that are compatible with standard complementary metal oxide semiconductor (CMOS) technologies. The resonant wavelength of Tamm absorbers is also highly tunable, and does not directly depend on the plasma frequency of the absorber material. Planar Tamm absorbers made of the same materials as the sensor discussed in this paper can be tuned to exhibit sharp resonant features corresponding to the Tamm state excitation at any wavelength across the optical and infrared range (supplementary Fig. S7). Furthermore, periodicity of the 1D photonic crystal is not required to achieve perfect absorption in Tamm structures, which further increases their tunability. As long as the conjugate-impedance matching condition is achieved, non-periodic Tamm absorbers can be designed and fabricated (Supplementary Fig. S5). Furthermore, as the Tamm interfacial states exist in both *p* and *s* polarizations, the Tamm absorbers can be tuned to operate with the *s*-polarized light excitation (Supplementary Fig. S5). Structure supporting multiple Tamm interfacial states at several frequencies can also be developed.

Even though the Tamm absorbers developed in this work were only characterized as temperature sensors, this system can be used for other sensing purposes including monitoring bio/chemical binding events on the surface. As the optical Tamm state is strongly confined at the interface between PhC and metal, the interaction of the enhanced electric field near the interface with the environment is weak. However, unlike SPP or LSP, the optical Tamm state is supported not only by the interface but by the whole structure that provides the conjugate-matching condition. Therefore, any changes of the refractive index in the environment change the optical impedance of the PhC $Z_{PhC}$, even though they happen far from the interface at which the optical Tamm state is supported. Thus, the change in the resonant condition of the optical Tamm state due to surface events can be detected by monitoring the ellipsometric angle Δ (Supplementary Fig. S8). In our absorber configuration, the field enhancement due to the optical Tamm state excitation occurs inside the multi-layered structure. However, planar absorbers can be designed to support interfacial optical Tamm states on their top air-matter interface (Supplementary Fig. S9). By carefully designing the structure, the electric field of the optical Tamm state can probe the events on the surface, and thus it is expected that highly sensitive optical sensing can be achieved. Therefore, we believe that the singular-phase detection with Tamm perfect absorbers will provide a new platform for a wide variety of sensing applications.

## Acknowledgements

This work was supported by DOE BES Grant No. DE-FG02-02ER45977. Y. T. is supported by the Funai Foundation for Information Technology through the Funai Overseas Scholarship.

## Methods

**Optical design.** The calculations of the reflectance, optical impedance and ellipsometric angles was done by the semi-analytical transfer matrix method. In the optical design of the Tamm structure, the reflection coefficients of the two structures that comprise the Tamm structure: a 1D PhC and a thin Au film on a-Si substrate are separately calculated to obtain the optical impedances $Z_{PhC}$ and $Z_{metal}$ that satisfy the complex conjugate matching $Z_{PhC} = Z_{metal}^*$. Then, the optical impedance of the Tamm structure that combines the two structures above is checked to provide free-space impedance matching. The reflectance and the ellipsometric angles are also calculated from the reflection coefficients obtained by the transfer matrix method, and temperature-dependence of them is included *via* the temperature-dependence of the dielectric function.

**Fabrication**. The alternative layers of $SiN_x$ and $SiO_2$ and the gold thin film were deposited with PECVD, and electron beam deposition, respectively. Figure 3a shows the SEM image of one of the fabricated samples (SEM 6010LA, JEOL USA, Inc). The buffered oxide etching was done to create topographical difference in SiN$_x$ and SiO$_2$ layers to obtain clear cross-section images. These layers were deposited on two types of substrates: single crystalline sapphire with the size of 10/10/0.5 mm (MTI corporation) and polycrystalline silicon. No adhesion layer between the gold thin film and the substrate was deposited. The measurement results of the fabricated samples with different substrates were similar. Therefore, the measurement results shown in the main text are all obtained from the sample with polycrystalline silicon substrate and those from the sample with crystalline sapphire substrate can be found in the supplementary information (supplementary Fig. S10).

**Reflectance amplitude and phase measurements.** Reflectance and ellipsometric angles were measured by a commercial ellipsometric spectrometer (M-2000, J.A. Woolam Co.). A simple schematic of the ellipsometer is shown in Fig. 3b. The wavelength resolution of the ellipsometer is 1.5nm, and the incident angle can be controlled by 0.01$^o$. In every set of measurements, the spot of the light incident on the sample is not at exactly the same position. Thus, after we set the incident angle of the light around 60$^o$ where the sample is expected to exhibit the reflectance minimum, the incident angle that guarantees the actual minimum reflectance for the spot on the sample was found. After we set the incident angle, the change of the ellipsometric angles was monitored as a function of the temperature. The temperature of the sample was controlled by a home-built temperature control system, which was designed to maintain a uniform temperature across the sample. This system is composed of Al bulk (2cm/2cm/5mm) on a ceramic heater on top of a glass cover. The glass cover at the bottom is to reduce the heat loss to the stage and the Al bulk is to keep the temperature uniform and stable. The temperature uniformity was checked by simulating the temperature distribution by the finite element method (COMSOL Multiphysics) with a given heat transfer coefficient to model the heat loss by convection and radiation.

**Supporting information**: Supplementary figures S1-S10.

Supplementary information

# Topological engineering of interfacial optical Tamm states for highly-sensitive near-singular-phase optical detection

Yoichiro Tsurimaki[1], Jonathan K. Tong[1], Victor N. Boriskin[2], Alexander Semenov[3], Mykola I. Ayzatsky[2], Yuri P. Machekhin[4], Gang Chen[1,*] and Svetlana V. Boriskina[1,*]

[1] Department of Mechanical Engineering, Massachusetts Institute of Technology, Cambridge, MA, USA

[2] National Scientific Center 'Kharkiv Institute of Physics and Technology', Kharkiv, Ukraine

[3] National Technical University 'Kharkiv Polytechnic Institute', Kharkiv, Ukraine

[4] Kharkiv National University of Radio Electronics, Kharkiv, Ukraine

emails: gchen2@mit.edu, sborisk@mit.edu

**Material parameters of dielectric and metal layers in the planar Tamm absorber**

Figure S1 (a) and (b) show the dielectric function of materials that comprise the Tamm structure in the main text. The room-temperature dielectric functions of amorphous Si, $SiO_2$, and $Si_3N_4$ are taken from Palik [1]. The temperature-dependent dielectric function of 50nm thin Au film is taken from Reddy et al. [2], where it was measured and modeled by Drude and 2 critical points model. The parameters in the model as a function of temperature for 50nm thin Au film were experimentally determined and listed in [2]. In our calculation, the curve fitting of the temperature dependence of the parameters was done to model the dielectric function of the 50nm thin Au film near ambient temperatures.

Figure S1 (c) and (d) show the variation of the dielectric function of the 50nm thin film in our model as a function of the temperature at $\lambda$=751.5nm, where the critical coupling to the optical Tamm states occurs in our simulation. Both real and imaginary parts of the dielectric function show the linear change near the room temperature. The slopes of the curves were estimated to be $d\text{Re}[\varepsilon]/dT$=-6.95·10$^{-3}$ K$^{-1}$, and $d\text{Im}[\varepsilon]/dT$= 6.103·10$^{-4}$ K$^{-1}$.

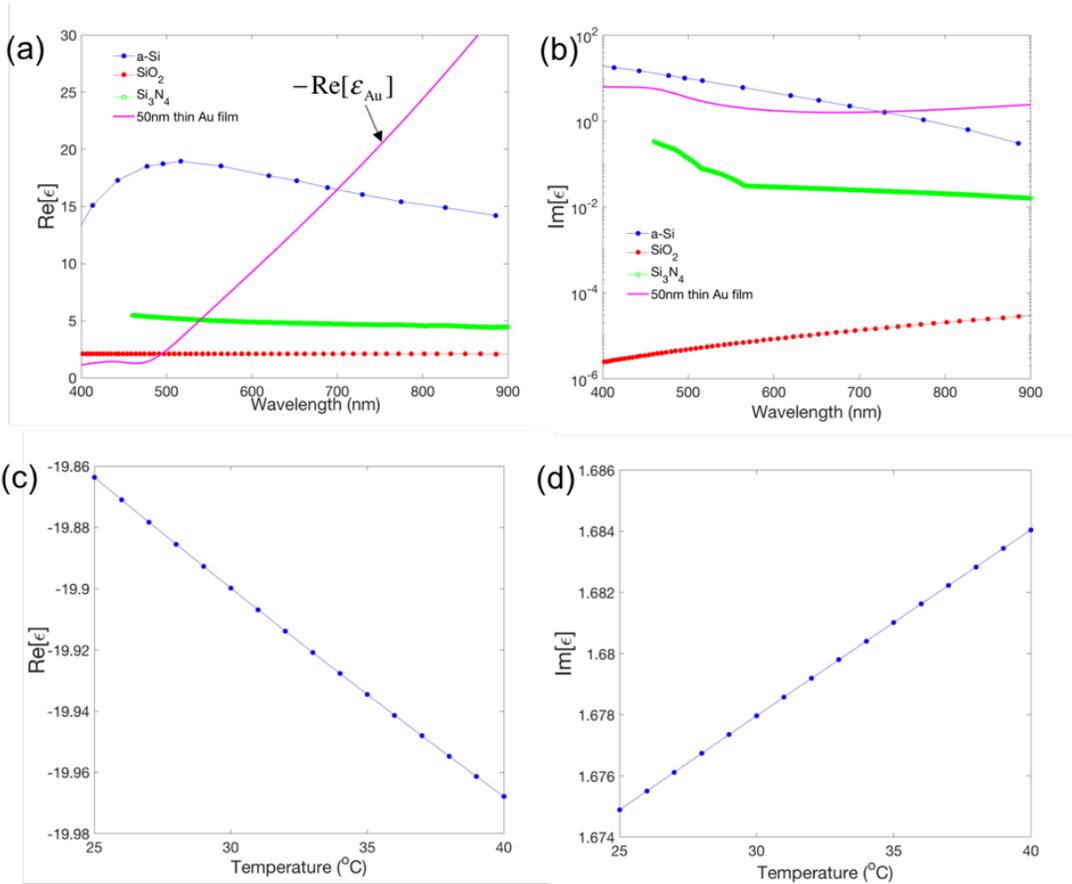

**Fig. S1 Dielectric functions of the materials comprising the Tamm structure and the temperature-dependent dielectric function of 50nm thin Au film at 751.5nm.** (a) real and (b) imaginary parts of the dielectric function of a-Si, SiO$_2$, Si$_3$N$_4$, and 50nm thin Au film. Data are taken from Palik [1] and Reddy et al [2]. (c) real and (d) imaginary parts of the dielectric function of 50nm Au film at temperatures from 25 to 40°C in our model.

## Dispersion relation of optical Tamm state

Figure S2 (a) shows the bandstructure of the 1D periodic photonic crystal made of 7 periods of $SiO_2$ (150nm) and $Si_3N_4$ (90nm). q and L are the Bloch wavevector and the period of the PhC, respectively. Also, Fig. S2 (b) shows the reflectance of the PhC (black line) and the metal-dielectric structure supporting the optical interfacial Tamm state (the PhC on top of 50nm thin gold film (no substrate in this structure)) (red line). For both figures, the incident angle of light is 49.2°, at which the critical coupling to the optical Tamm state occurs in this geometry of the absorber. As shown in Fig. S2 (a), the optical interfacial Tamm state exists in the first photonic bandgap of the 1D periodic PhC. Around the Tamm resonant frequency, the reflectance of the structure is high due to the high reflectance of the PhC. For longer wavelengths than the photonic bandgap of the PhC, the reflectance of the structure is high, as the thin gold film becomes more reflective. For shorter wavelength, the thin gold film is no longer reflective due to the direct interband absorption in Au, and the reflectance of the Tamm structure follows that of the PhC without the thin gold film. Figure S2 (c) and (d) show the dispersion relations of the optical interfacial Tamm states for *p-* and *s-*polarized light, respectively. The change of the incident angle from the normal corresponds to change the horizontal wavevector that is tangential to the interfaces of PhC and it is constant throughout the PhC due to the boundary condition for Maxwell equations. The scale bar shows the reflectance values. The dispersion of the Tamm interfacial state corresponds to the loci of low reflectance around 2000 – 2500 $10^{12}$[rad/s]. As shown in the figure, the dispersion is parabolic for both polarizations with slightly different 'effective mass'. The optical Tamm state can be excited by light incident at angles ranging from 0° (i.e., normally to the structure) to 80° for *p-*polarization, and from 0° to around 70° for *s-*polarization. The optical interfacial Tamm state is always in the photonic bandgap of the 1D PhC, and thus it shows the sharp and deep reflectance peak. The dispersion relation of the Tamm state can also be obtained by numerically conjugate matching the optical impedances of the two composite media on the opposite sides of the material interface.

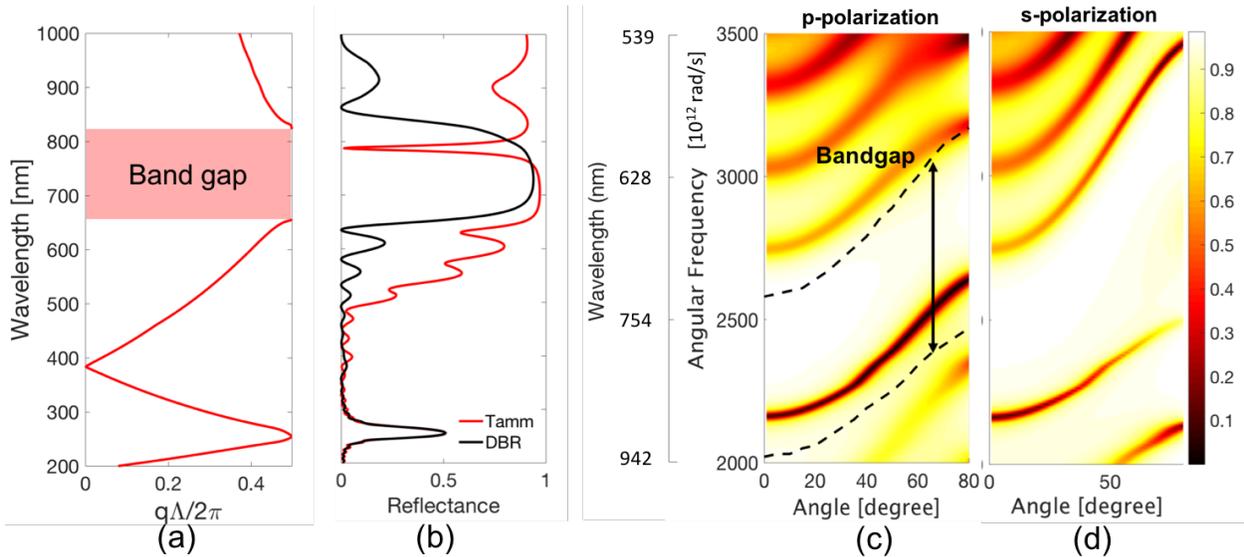

**Fig. S2 Characteristics of optical Tamm state.** (a) Bandstructure of 1D periodic PhC (7periods of $SiO_2$ (150nm thick) and $Si_3N_4$ (90nm thick)). (b) Reflectance of the 1D PhC (black line, without thin gold film) and of the Tamm structure (red line, PhC and the thin gold film (50nm thick)). (c), (d) Dispersion relations of *p-* and *s-*polarized optical interfacial Tamm states. The scale bar shows the values of the reflectance of the structure.

# Tamm resonances in the angular spectra of conjugate-matched planar absorbers

Figure S3 shows the angular dependence of the reflectance of the conjugated matched Tamm absorber in Fig. 1. The reflectance for both polarizations calculated at the resonant wavelength of 751.5nm is shown. As seen, the *p*-polarized reflectance shows a sharp reflectance dip at the incident angle of 60 degrees to normal due to the critical coupling to the optical Tamm state. Tamm absorbers can be exploited as optical sensors that monitor the shifts of the angular reflectance spectra acquired at a fixed wavelength.

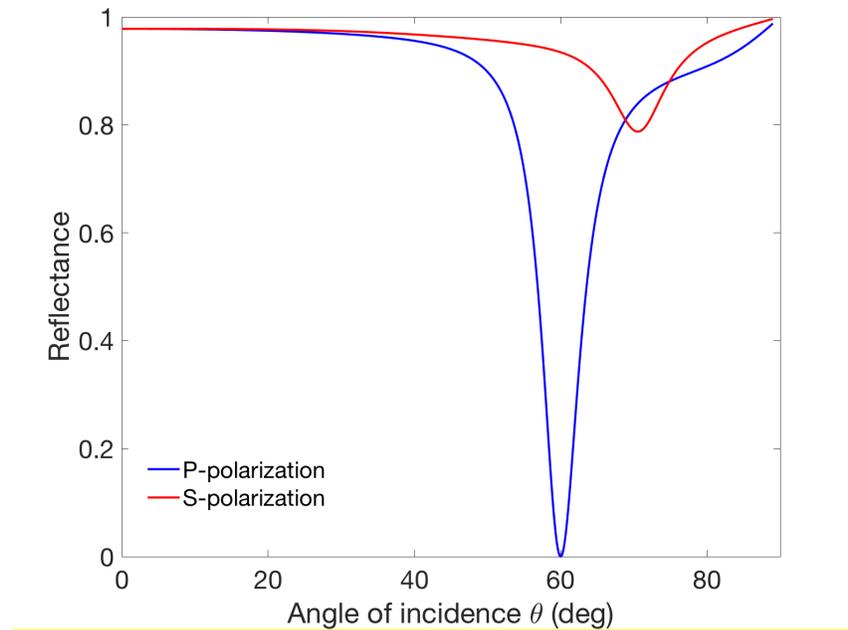

**Fig. S3 Angular reflectance spectrum of the Tamm absorber in Fig. 1a**

**for both polarizations calculated at 751.5nm.**

# Determination of a spectral width and a Q-factor of a resonant feature

The spectral widths of the reflectance and the ellipsometric angles $\Psi$ and $\Delta$ were determined by the curve fitting to the calculated and measured data. The parameters for the curve fitting can be found in Table S1 below. Figures S4a,b show the *p*-polarized reflectance spectra obtained via simulation (Fig. 1c) and experiment (Fig. 4c). The curve fitting to them was obtained by using the Lorentz distribution shown in Eq. S1:

$$R = R_\infty \left( 1 - \left(1 - \frac{R_{min}}{R_\infty}\right) \frac{\left(\frac{\Gamma}{2}\right)^2}{(\omega - \omega_0)^2 + \left(\frac{\Gamma}{2}\right)^2} \right), \quad \text{(S1)}$$

where $R_{min}$ and $R_\infty$ are the reflectance at the resonant frequency, and at a frequency far enough from the resonance, respectively. The line widths for the spectral features observed in Fig. S4 (a) and (b) were calculated to measure 18nm and 15nm, respectively.

Figure S4 (c) and (d) show the simulated (Fig. 2) and measured (Fig. 4e) frequency spectra of the ellipsometric angle $\Psi$. The curve fitting to them was obtained by using the Lorentz distribution shown in Eq. S2:

$$\Psi = \Psi_\infty \left( 1 - \left(1 - \frac{\Psi_{min}}{\Psi_\infty}\right) \frac{\left(\frac{\Gamma}{2}\right)^2}{(\omega - \omega_0)^2 + \left(\frac{\Gamma}{2}\right)^2} \right), \quad \text{(S2)}$$

where $\Psi_{min}$ and $\Psi_\infty$ are the ellipsometric angles at the resonant frequency, and at a frequency far enough from the resonance. As a result, the spectral widths were found to be 8.3nm (Fig. S4c), and 6.9nm (Fig.S4d).

Figure S4e shows the simulated ellipsometric angle $\Delta$ and the curve fitting to it by the following equation:

$$|\cos\Delta| = a\frac{\omega - \omega_1}{\omega_1} + 1 - \frac{r^2(\omega - \omega_2)^2 + (1-r^2)(\Gamma/2)^2 - 2r\sqrt{1-r^2}(\omega - \omega_2)\Gamma/2}{(\omega - \omega_2)^2 + (\Gamma/2)^2} \quad \text{(S3)}$$

where $a$ is a constant, $\omega_1$, $\omega_2$ are frequencies, $r$ is the amplitude of reflection coefficient, and $\Gamma$ is the spectral width. In the curve fitting, we chose $\omega_1=\omega_2=\omega_0$, which is the resonant frequency, and $a$, $r$, and $\Gamma$ were fitting parameters. The parameters used for the best fitting are listed in Table S1. As a result, we found that the spectral width was 0.006nm. This spectral width corresponds to the wavelength range in which the asymmetric spectral shape of $|\cos\Delta|$ varies from its minimum to maximum. In order to determine the spectral width of the measured ellipsometric angle $\Delta$ in Fig. 4e, we designed the structure that shows a similar $\Psi_{min}$ to the experimentally measured $\Psi_{min}$=1.78 at the same wavelength 744.2nm. The structure is the same as in Fig. 1a except for the incident angle, and the thickness in the last layer of $Si_3N_4$, which were chosen to be 63° and 87.6nm, respectively. This structure shows $\Psi_{min}$=1.872 at 744.2nm, which is close to the experimental data in Fig. 4e. Then, we conducted the curve fitting to obtain the spectral width by using Eq. S3 shown above. As a result, we found that the spectral width is 0.52nm, which corresponds to the wavelength range of the peak-to-peak variation of $|cos\Delta|$ in Fig. S4f. We expect that the measured $\Delta$ in Fig. 4e also shows a similar variation around that resonant wavelength, which we missed due to the low spectral resolution of the ellipsometer. Therefore, we used this spectral width to estimate the figure of merit in the main text.

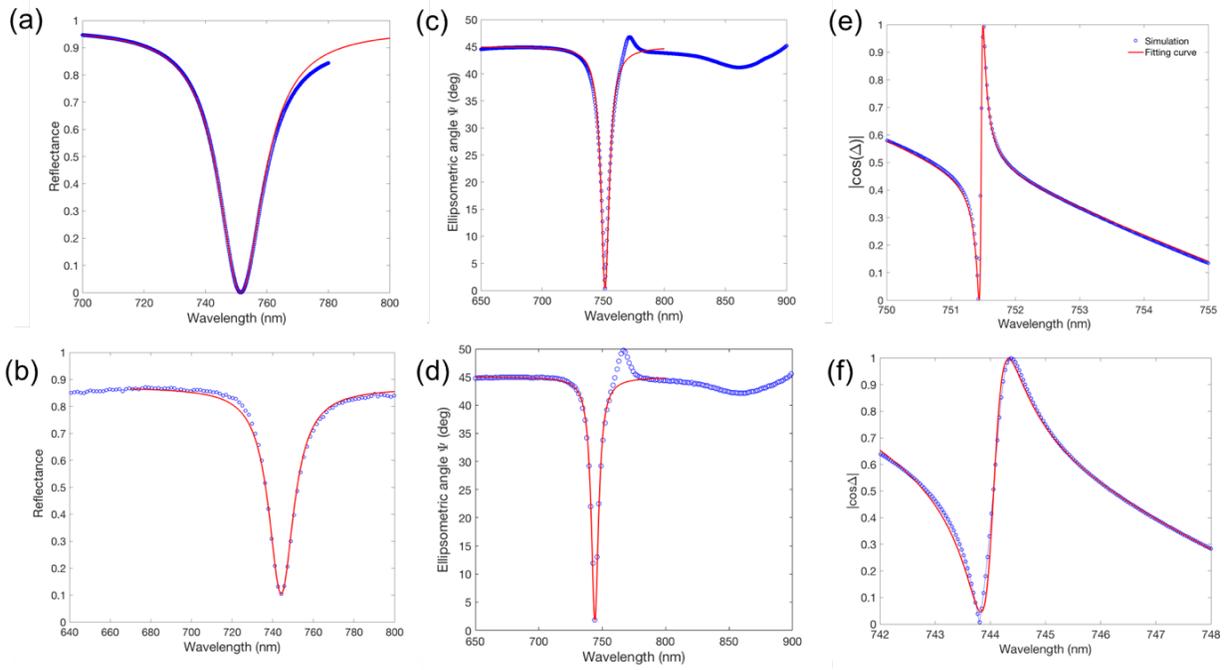

**Fig. S4 Extraction of the spectral width of the reflectance and ellipsometric angles by curve fitting**. (a) Simulated (Figs. 1c) and (b) experimental spectra (Fig. 4c) of *p*-polarized reflectance. (c) Simulated (Fig. 2) and (d) experimental spectra (Fig. 4d) of the ellipsometric angle $\Psi$. (e) Simulated spectrum of the ellipsometric angle $\Delta$ in Fig. 2. (f) Simulated spectrum of the ellipsometric angle $\Delta$ under the condition that exhibits the same $\Psi_{min}$ at the same wavelength as the experimental result in Fig. 4e.

Table S1: Fitting parameters of Eq. S1, Eq. S2, and Eq. S3

| Name | Figure | $\omega_0$ [rad/s] | $\Gamma/\omega_0$ | Q-factor |
|---|---|---|---|---|
| Reflectance | Fig. S6 (a) | 2.508 10$^{15}$ | 0.0235 | 40 |
|  | Fig. S6 (b) | 2.533 10$^{15}$ | 0.0202 | 49.5 |
| Elipsometric angle $\Psi$ | Fig. S6 (c) | 2.508 10$^{15}$ | 0.011 | 90 |
|  | Fig. S6 (d) | 2.533 10$^{15}$ | 0.0093 | 107 |

| Name | Figure | $\omega_1$ [rad/s] | $\omega_2$ [rad/s] | $\Gamma/\omega_0$ | r | a |
|---|---|---|---|---|---|---|
| Ellipsometric angle $\Delta$ | Fig. S6 (e) | 2.508 10$^{15}$ | 2.508 10$^{15}$ | 0.00008 | 0.685 | 70 |
| | Fig. S6 (f) | 2.5325 10$^{15}$ | 2.5334 10$^{15}$ | 0.0007 | 0.75 | 70 |

## s-polarized optical Tamm state supported by non-periodic structure with low-loss Ag absorber layer

Periodicity of the 1D photonic crystal is not required to achieve perfect absorption in Tamm structures, which further increases their tunability. As long as the conjugate-impedance matching condition is achieved, non-periodic Tamm absorbers can be designed. Metals other than gold can be used as an absorber layer, including silver, which has lower dissipative losses than gold in the visible frequency range. Furthermore, as the Tamm interfacial states exist in both *p* and *s* polarizations, the Tamm absorbers can be tuned to operate with the s-polarized light excitation. Figure S5 shows a multilayer structure composed of zinc sulfide (ZnS) layers alternated by magnesium fluoride (MgF$_2$) layers grown on top of a single Ag layer and a glass substrate. We designed these structures to exhibit a Tamm resonance at the wavelength of 633nm and for the *s*-polarized incident light. Samples were prepared by thermal evaporation technique with in-situ ellipsometry used to control the growth process, and their reflection characteristics were measured as a function of the illumination angle for both *p*- and *s*-polarized light excitation. The source of monochromatic light was a He–Ne laser of wavelength 6328 Å.

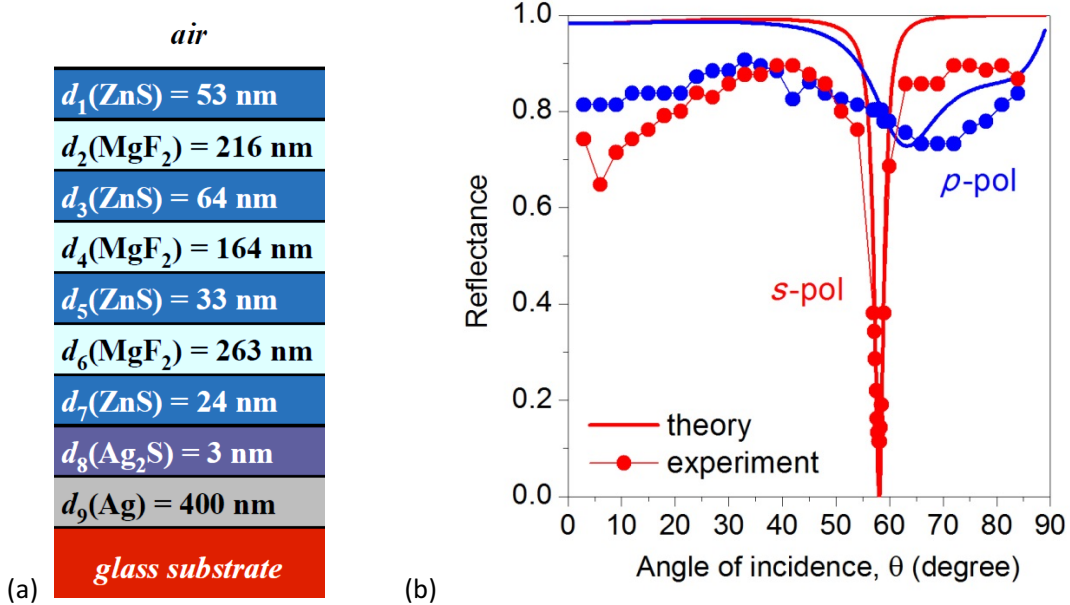

**Fig. S5.** *s*-polarized optical Tamm state supported by non-periodic structure with low-loss Ag absorber layer. (a) A schematic of the non-periodic Tamm structure composed of the dielectric thin films on top of a thick Ag layer. (b) Calculated (thick lines) and experimentally measured (thin lines with circles) angular reflectance spectra of the structure in (a) under illumination of *s*- and *p*-polarized light of a He–Ne laser.

## *p*-polarized optical Tamm state supported by a periodic structure with a CMOS-compatible Al absorber layer

The optical Tamm state can be realized by a variety of other materials including dielectrics, semiconductors and metals. In the main text, we used the thin film gold as one part of the structure that supports the optical Tamm state. However, the optical Tamm state can be supported by metals that are compatible with standard complementary metal oxide semiconductor (CMOS) technologies. Figure S6 shows the measured reflectance of the structure comprised of 4 periods of $SiO_2$ (150nm thick) and $Si_3N_4$ (90nm thick) on top of optically-thick Al layer (100nm). The incident angle is normal, and the temperature of the sample is 23$^o$. As seen in the figure, the reflectance shows a dip at around 850nm. In Fig. S6, the calculated reflectance is also shown and it shows a reasonable agreement with the measured reflectance. Therefore, it is concluded that the measured reflection minimum is due to the coupling to the optical Tamm state. The magnitude of the reflectance is different from the simulated reflectance. It is considered that this difference is due to sample quality and the calibration of the spectrometer.

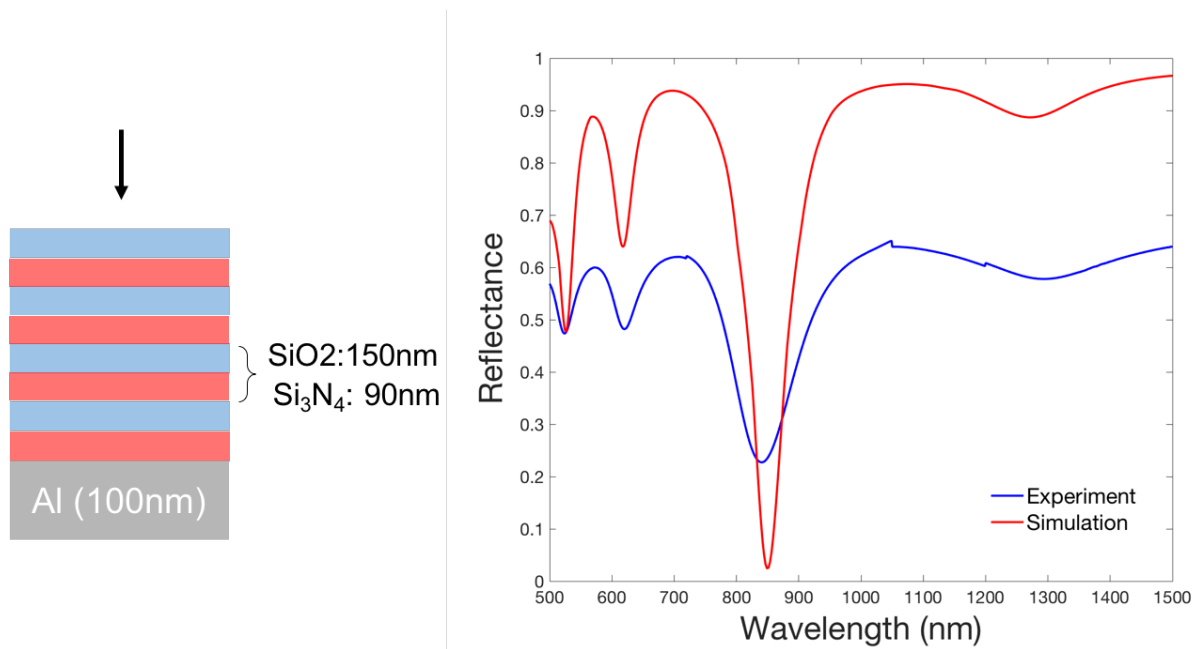

**Fig. S6. *p*-polarized optical Tamm state supported by a periodic structure with a CMOS-compatible Al absorber layer.** Reflectance of the structure comprised of 4 periods of $SiO_2$(150nm) and $Si_3N_4$(90nm) on top of optically-thick Al layer (100nm). Incident angle is normal and the temperature of the sample is 23$^o$.

# Spectral tunability of planar Tamm absorbers

Planar multi-layered absorbers can be designed to exhibit critical coupling to the optical Tamm states in a wide range of wavelengths. As an example, three structures showing the resonant frequencies in different spectrum region, i.e. visible (653nm) and near-infrared (911nm and 1.4µm), for the normal incident light are designed as shown in Fig. S7. All the structures are composed of 7 periods of PhC made of $SiO_2$ and $Si_3N_4$ on top of 50nm Au film and an amorphous silicon substrate. The resonant frequencies are tuned only by varying the thicknesses of the PhC layers. The PhC layer thicknesses of the three structures are listed in Table. S2 below.

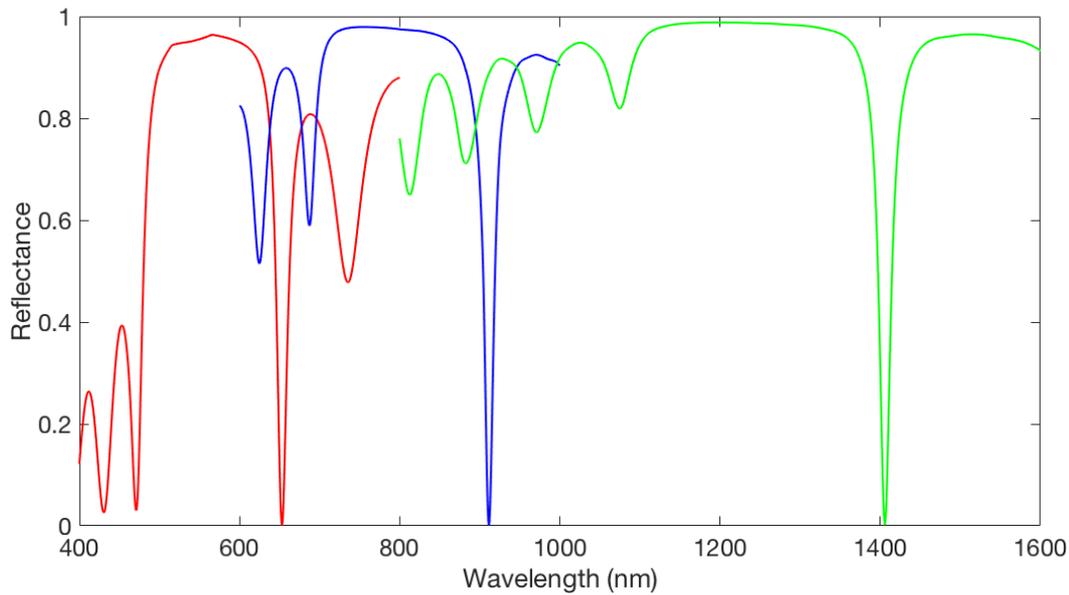

**Fig. S7. Tamm structures with different PhC thicknesses showing resonant frequencies in visible (653nm), and near infrared (911nm and 1.4µm).**

Table S2: PhC layer thicknesses of three Tamm absorbers in Fig. S7

| Line color in Fig. S7 | Resonant wavelength (nm) | SiO2 (nm) | $Si_3N_4$ (nm) | Last layer of $Si_3N_4$ (nm) |
|---|---|---|---|---|
| Red | 653 | 100 | 60 | 80 |
| Blue | 911 | 150 | 90 | 110 |
| Green | 1406 | 225 | 155 | 175 |

## Tamm sensor as label-free bio/chemical sensor

Label-free detection of bio/chemical events on the surface needs to be both sensitive and specific to the target material. The sensitivity depends on the device and its transduction mechanism, while the specificity and the affinity to the target is determined by the functionalization of the surface that is provided through chemical modification of the surface.

Figure S8a shows the simulated results of the change of the ellipsometric angle $\Delta$ in response to the change of the refractive index in the environment. The structure is the same as the optimally-designed one shown in Fig. 1a, the incident angle is $60^\circ$ and the wavelength is at 751.5nm at which the critical coupling to the optical Tamm state is attained. At this wavelength, $\Psi_{min}$ ~ 0.2. As shown in Fig. S8a, the ellipsometric angle $\Delta$ shows a linear variation as the refractive index in the environment increases. The sensitivity slope is calculated as $d\Delta/dn=10997$ (deg·RIU$^{-1}$). Figure S8 (b) shows the simulated results of the change of the ellipsometric angle $\Delta$ as a function of the refractive index in the environment for the as-fabricated structure with the optical characteristics as in Fig. S4f, which is our modeling of the experimental results in Fig. 4e. In this case, the slop is estimated to be $d\Delta/dn=1043$ (deg·RIU$^{-1}$). Both Figure S8a and b show that the Tamm absorber can detect the change of the refractive index in the environment. Figure S8c shows the simulated results of the change of the ellipsometric angle $\Delta$ when a layer with different refractive index exists at the surface. The layer thickness is chosen to be 1nm and 2nm to model adsorbate molecules on the surface. The structure, incident angle, and the wavelength are the same as those used in Fig.1a. The red and blue dots in Fig. S8c show the cases for the surface layer thickness of 1nm and 2nm, respectively, and the broken lines show the linear fitting to the dots. It is found that the slope is $d\Delta/dn=52$ (deg·RIU$^{-1}$), and $d\Delta/dn=105$ (deg·RIU$^{-1}$), for the layer thickness of 1nm and 2nm, respectively. Therefore, we showed simulation results that the Tamm absorbers can be used to monitor bio/chemical events on the surface.

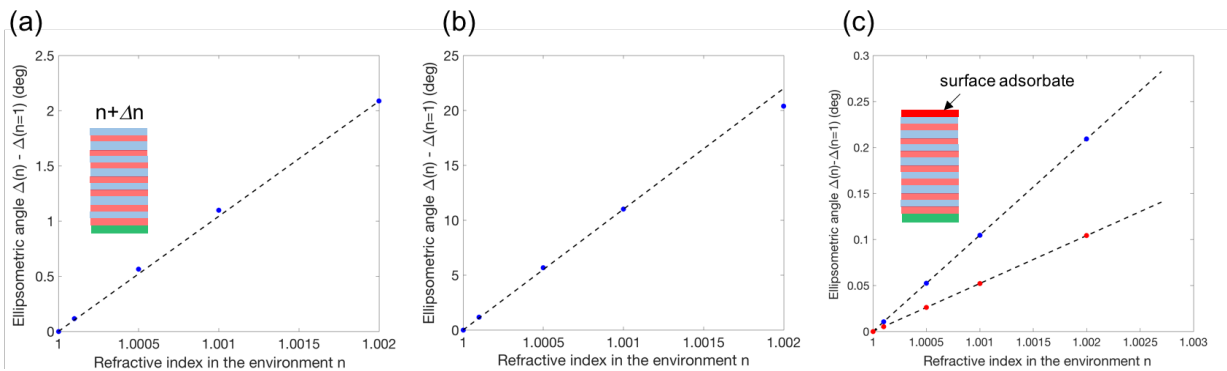

**Fig. S8. Tamm absorbers as bio/chemical sensors.** (a) and (b) Change of ellipsometric angle $\Delta$ as a function of the change in refractive index in the environment for the ideal designed structure in Fig. 1a, and for the as-fabricated structure with the optical spectra as in Fig.S4f, respectively. Blue dots are simulated results and broken lines are linear fitting to the simulation. (c) Change of ellipsometric angle $\Delta$ under the existence of a surface layer with different thickness and different refractive index. The condition is the same as in Fig.1a. The red and blue dots are simulated results for the surface layer thickness of 1nm and 2nm, respectively and the broken lines are linear fitting to the simulation.

## Coupling to the optical Tamm state from the metal side

The Tamm absorber can be designed so that the incident light from the metal side is perfectly absorbed at a single narrowband frequency at a specific angle. Figure S9 shows the simulation result of p-polarized reflectance as a function of the wavelength for the structure comprised of 30nm Au film and 9 periods of the PhC made of $SiO_2$ (150nm) and $Si_3N_4$ (90nm). The thickness of $Si_3N_4$ layer beneath the Au layer is tuned to 89nm to attain the perfect absorption. The incident angle is 60 degrees. The same dielectric function of 50nm Au film is used for the 30nm Au film.

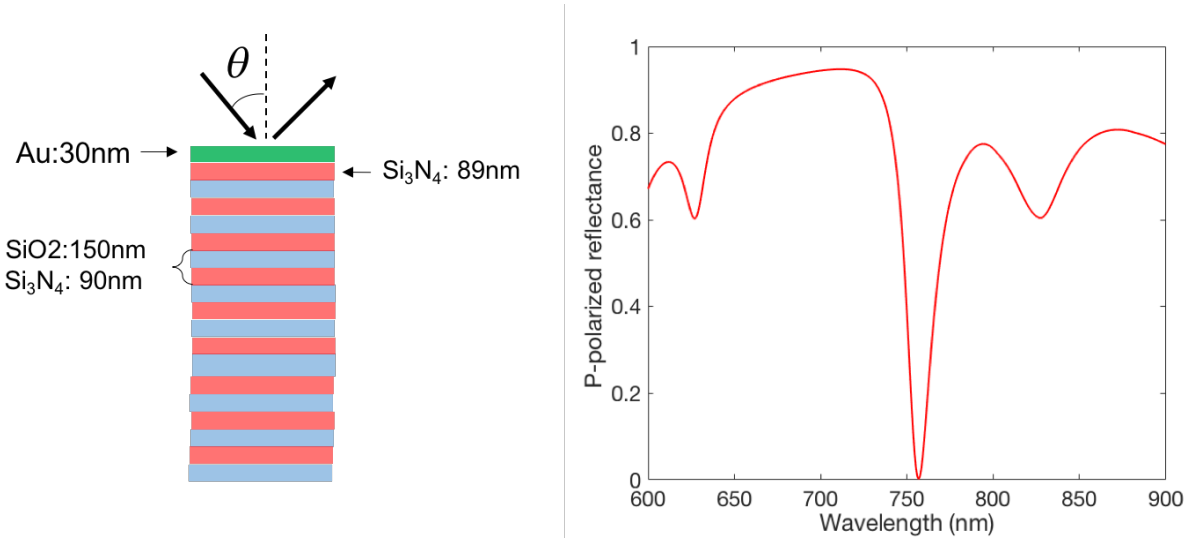

**Fig. S9. Tamm absorbers for the incident light from the metal side.** Reflectance of the structure comprised of 30nm Au film and 9 periods of $SiO_2$(150nm) and $Si_3N_4$(90nm) with the last layer thickness of $Si_3N_4$ 89nm. The incident light is p-polarized and the incident angle is 60 degrees.

## Singular-phase detection with a planar Tamm absorber on a sapphire substrate

The singular phase detection by using the same structure as in the main text (7 periods of $SiO_2$ (150nm) and SiN (90nm) as well as the Au layer (50nm)) on top of single-crystalline sapphire substrate is discussed in this section. The incident angle is 55°. Figures S10a,d show the ellipsometric angles $\Psi$ and $\Delta$ as a function of the wavelength at 25°C. The coupling of the incident light into the optical Tamm state occurs at around 782nm. The minimum of the ellipsometric angle $\Psi_{min}$ was around 4.8°. Figures S10b,c show the temperature dependence of the ellipsometric angle $\Psi$. As seen in the figure, the ellipsometric angle $\Psi$ varies linearly as the temperature changes. From the slope of the variation of the ellipsometric angle $\Psi$, the sensitivity was found $d\Psi/dT=0.12K^{-1}$. The sample with the sapphire substrate showed the same sensitivity as the sample in the main text for the amplitude sensing. However, the sample with the sapphire substrate showed a larger $\Psi_{min}$~4.8° than the sample discussed in the main text ($\Psi_{min}$~1.8°). Thus, according to the scaling law in Fig. 6 in the main text, the sample with the sapphire substrate would show lower sensitivity for the phase sensing, i.e. lower $d\Delta/dT$. The temperature dependence of the ellipsometric angle $\Delta$ of the sample with sapphire substrate is shown in Figs.

S10e,f. In fact, the sensitivity is found to be $d\Delta/dT \sim 1.3°$, smaller than the sensitivity exhibited by the sample in the main text. However, note that the relation between $\Psi_{min}$ and $d\Delta/dT$ also follows the scaling law in Fig. 6 in the main text. Therefore, irrespective of the type of substrate, one should be able design a structure that attains smaller $\Psi_{min}$ in order to attain the higher sensitivity in singular-phase sensing.

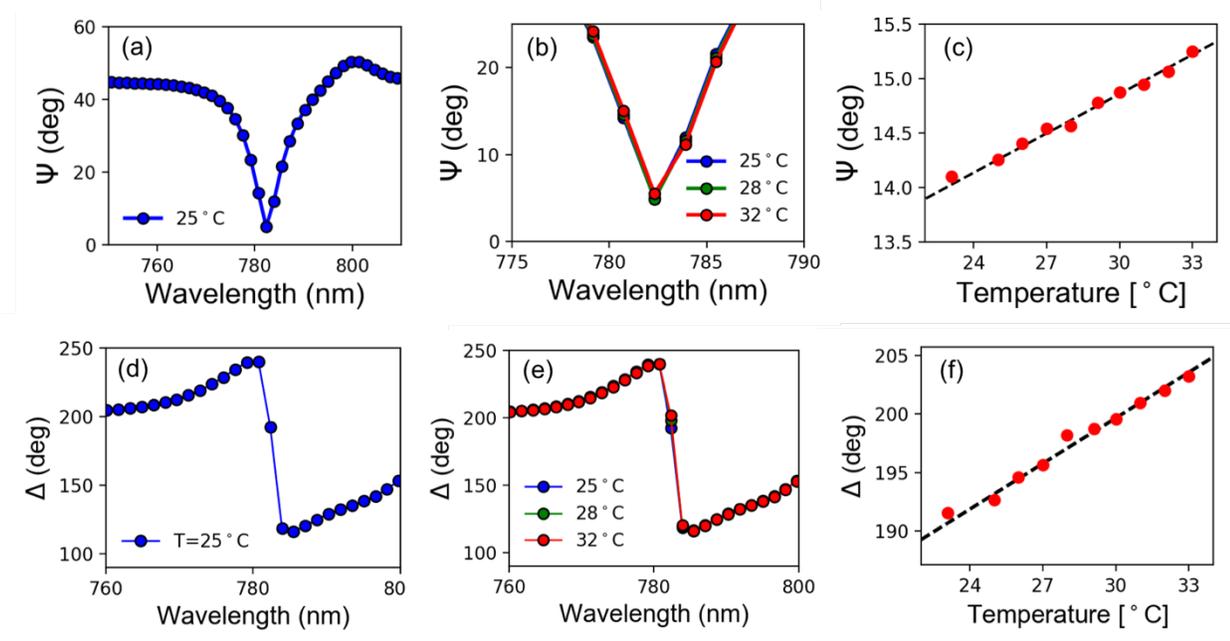

**Fig. S10. Measured ellipsometric angles at different temperatures.** (a) Measured ellipsometric angle $\Psi$ as a function of wavelength at room temperature. (b) Ellipsometric angle $\Psi$ as a function of the wavelength at different temperatures. (c) Ellipsometric angle $\Psi$ as a function of the temperature. (d) Measured ellipsometric angle $\Delta$ as a function of wavelength at room temperature. (d) Ellipsometric angle $\Delta$ as a function of the wavelength at different temperatures, and (f) as a function of the temperature.